\documentclass[prb,twocolumn,superscriptaddress,showpacs,amsmath,amssymb]{revtex4-1}

\newcommand{\dd}[1]{\mathrm{d}#1\,}

\renewcommand{\Re}{\mathop{\mathrm{Re}}}
\renewcommand{\Im}{\mathop{\mathrm{Im}}}
\newcommand{\Tr}{\mathop{\mathrm{Tr}}}

\newcommand{\tr}{\mathop{\mathrm{tr}}}
\newcommand{\sgn}{\mathop{\mathrm{sgn}}}

\renewcommand{\vec}[1]{\bm{#1}}

\usepackage{graphicx}
\usepackage{latexsym}
\usepackage{amsmath}
\usepackage{amssymb}
\usepackage{amsfonts}
\usepackage{color}
\usepackage{units}
\usepackage{bm}
\usepackage{verbatim}
\usepackage{hyperref}

\begin{document}

\title{Stimulated quasiparticles in spin-split superconductors}

\author{P.~Virtanen}
\affiliation{Low Temperature Laboratory, Department of Applied Physics, P.O. Box 15100, FI-00076 Aalto University, Finland}
\affiliation{University of Jyvaskyla, Department of Physics and Nanoscience Center, P.O. Box 35, 40014 University of Jyv\"askyl\"a, FINLAND}

\author{T.T.~Heikkil\"a}
\affiliation{University of Jyvaskyla, Department of Physics and Nanoscience Center, P.O. Box 35, 40014 University of Jyv\"askyl\"a, FINLAND}

\author{F.S.~Bergeret}
\affiliation{ Centro de F\'{i}sica de Materiales (CFM-MPC), Centro
Mixto CSIC-UPV/EHU, Manuel de Lardizabal 5, E-20018 San
Sebasti\'{a}n, Spain}
\affiliation{Donostia International Physics Center (DIPC), Manuel
de Lardizabal 5, E-20018 San Sebasti\'{a}n, Spain}

\begin{abstract}
  In superconductors spin-split by an exchange field, thermal effects
  are coupled to spin transport. We show how an oscillating
  electromagnetic field in such systems creates spin imbalance, that
  can be detected with a spin-polarized probe. The sign and
  magnitude of the probe signal result from a competition between processes converting
  field induced spin energy imbalance to spin imbalance, dominant at
  low frequencies, and
  microwave-driven pair breaking at high frequencies. In the presence of spin-flip scattering,
  we show that ac excitation also leads to multistabilities in 
  the superconducting state.
\end{abstract}

\maketitle

\section{Introduction}

\begin{figure}
  \includegraphics{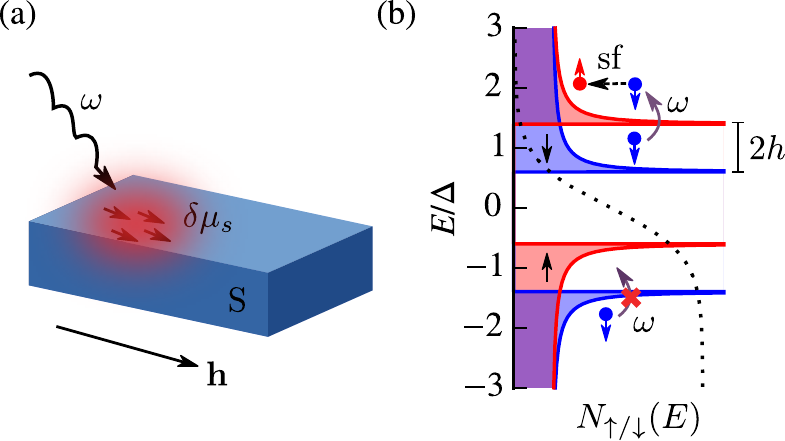}
  \caption{
    \label{fig:schematic}
    (a)
    A superconductor absorbs microwave electromagnetic radiation, in the
    presence of an internal exchange field $\vec{h}$. In the steady
    state, this generates a
    spin imbalance $\delta\mu_s$, an excess of quasiparticles
    whose spins are either aligned ($\delta\mu_s>0$) or anti-aligned
    ($\delta\mu_s<0$) with the axis of the exchange field.
    (b)
    The quasiparticle spectrum is spin-split by the exchange field. 
    Coupling to microwaves generates spin-conserving quasiparticle 
    transitions that change energy by $\pm\omega$
    and perturb the electron distribution (dotted).
    Elastic spin-flip scattering transforms quasiparticles
    to the opposite spin species, converting   energy imbalance
    to spin imbalance.
  }
\end{figure}

Long-lived spin excitations are interesting for spintronics
applications, and the spin transport in superconductors has
recently attracted renewed attention in this context.
\cite{linder2015-ss,kivelson1990-bqs,zhao1995-trs} 
Spin accumulation
in a superconductor can be generated by injecting current from a
spin-polarized electrode, for example a ferromagnet. A second approach
for spin injection studied in a number of recent experiments
\cite{quay2013-sia,huebler2012-lrs,wolf2013-sin,wolf2014-spq,quay2014-fdm,quay2015-qsr}
is to use a magnetic field or proximity to ferromagnetic insulators to
Zeeman split the density of states of the superconductor
\cite{tedrow1971-sdt,tedrow1986-spe} (cf. Fig.~\ref{fig:schematic}),
so that injection of current from an unpolarized probe also generates
observable spin accumulation. The spin-splitting also changes the
quasiparticle physics so that one component of the imbalance only
relaxes via inelastic scattering,
\cite{huebler2012-lrs,quay2013-sia,silaev2015-lrs,bobkova2015-lrs}
leading to long observed spin lifetimes and relaxation lengths.

The physics of the long-ranged quasiparticle spin accumulation in
spin-split superconductors is closely connected to their
thermoelectric properties.  Magnetic interactions in superconductors
break the spin-resolved electron-hole symmetry, enabling large
thermoelectric responses.  This is predicted to occur due to magnetic
impurities \cite{kalenkov2012-tlt}, spin-active interfaces,
\cite{machon2013-nte,kalenkov2014-ehi,machon2014-gte} and in
superconductor--ferromagnet systems in the presence of exchange
fields. \cite{ozaeta2014-hte} The large thermoelectric effect in spin-split superconductor/ferromagnet
tunnel junction has been  observed very recently (see
Ref.~\onlinecite{kolenda2015-otc}).

The thermoelectric mechanisms are also connected to photoelectric
effects in superconductors \cite{zaitsev1986-pes,kalenkov2015-dsb},
where absorbed radiation is converted to a dc voltage observed in a
probe electrode.  Based on the above discussion, a photo-spin-electric effect should be present also in spin-split
superconductors --- the absorbed radiation generates spin imbalance,
which relaxes slowly via inelastic scattering. This is interesting to
consider e.g. in the context of measurements that use microwave signals to
probe spin resonances of the quasiparticles. \cite{quay2015-qsr}

In this work, we discuss how an electric ac field in diffusive spin-split
superconductors produces spin imbalance [see
Fig.~\ref{fig:schematic}(a)]. We find that the ac driving generates
spin imbalance that is either parallel or antiparallel to the exchange
field, depending on the drive frequency.  Nonequilibrium states
generated by ac fields in conventional superconductors have long been
studied, \cite{eliashberg1970-fss} and we extend the picture to
include spin splitting. Although interaction with the fields conserves
spin, combining it with elastic spin-flip scattering from
e.g. magnetic impurities results to a nonequilibrium steady state
with nonzero spin imbalance [see Fig.~\ref{fig:schematic}(b)]. We
discuss how the effect can be detected via ferromagnetic probes
[Eq.~\eqref{eq:Idetapprox}].  Similar photoelectric effects are known
to occur also in the absence of spin splitting,
\cite{zaitsev1986-pes,kalenkov2015-dsb} but they require weak elastic scattering. We also predict that the spin imbalance
results to an instability in the superconducting order parameter,
permitting multiple non-zero values for it in a temperature range
around $T_c$, leading to hysteresis and 
providing a second characteristic signature of the effect.

The manuscript is organized as follows.  We outline the model in
Sec.~\ref{sec:noneq} and discuss observables accessible with
spin-polarized electrical probes in Sec.~\ref{sec:probes}.
Modification of the superconducting order parameter is discussed in
Sec.~\ref{sec:gapeqn}, and we conclude in Sec.~\ref{sec:discussion}.

\section{Kinetic equations}
\label{sec:noneq}

We consider a diffusive superconductor film with a Zeeman field induced either by an external magnetic
field \cite{tedrow1971-sdt} or for example proximity to a magnetic
insulator.\cite{miao2015-smm,tedrow1986-spe}  In order to describe a nonequilibrium situation 
we apply the quasiclassical Keldysh-Green function
formulation, \cite{bergeret2005-ots,usadel1970-gde,kopnin2001-ton,morten2004-std,morten2005-stm} 
and write the Usadel equation for the spin-split superconductor (here and below, we set $\hbar=e=k_B=1$):
\begin{gather}
  \label{eq:usadel}
  D\hat{\nabla}\cdot(\check{g}\hat{\nabla}\check{g})
  +
  [i \epsilon\hat{\tau}_3 - i (\vec{h}\cdot\vec{S})\hat{\tau}_3 - \hat{\Delta} - i\check{\sigma}, \check{g}]
  =
  0
  \,.
\end{gather}
The function
$\check{g}(t,t')$ is a  matrix which in the Keldysh-Nambu-spin space has the form
\[
\check{g}=\left( \begin{array}{cc}
\hat g^R & \hat g^K\\
0 & \hat g^A
\end{array} \right)\; ,
\]
where $\hat g^{R,A,K}$ are the retarded, advanced and Keldysh 2$\times$2 
matrices in the Nambu ($\tau_j$)  and spin ($s_j$) spaces and   $\vec{S}=(s_1,s_2,s_3)$.
The exchange field $\vec{h}$
is induced by an external magnetic field \cite{tedrow1971-sdt} or for
example proximity to a magnetic insulator.
\cite{miao2015-smm,tedrow1986-spe} Here, $D$ is the diffusion
constant of the superconductor, $\Delta$ is the order parameter, and
$\check{\sigma}$ a self-energy corresponding to spin-flip and
inelastic scattering (electron-phonon or electron-electron). We use a
gauge where the electric potential is $\varphi=0$, and coupling to
electromagnetic fields is via a vector potential $\vec{A}$ appearing
in the covariant derivative
$\hat{\nabla}X = \nabla X - [i \vec{A}\tau_3, X]$.  

We assume that the superconducting film is in a uniform time-dependent
electric field ${\cal E}(t)=A_0\omega_0\sin(\omega_0 t)$.  We follow a
similar approximation procedure as in
Ref.~\onlinecite{eliashberg1970-fss}. In a spatially uniform
situation, assuming the film is thinner than the skin depth, the
vector potential enters equivalently as a self-energy
$\check{\sigma}_A(t,t')=-i D {\vec A}(t)\cdot \hat{\tau}_3
\check{g}(t,t') {\vec A}(t') \hat{\tau}_3$ which 
after time averaging is given by
\[
\check{\sigma}_A(E)=-i\frac{DA_0^2}{4}\tau_3[\check{g}(E+\omega_0)+\check{g}(E-\omega_0)]\tau_3\; .
\]
Considering time dynamics implied by Eq.~\eqref{eq:usadel}, this expression describes the effect of the ac field
 in the leading order in the small parameter $DA_0^2\ll\omega_0$.

The self-energy term in Eq. (\ref{eq:usadel}) also takes into account
 a number of relaxation processes present in real superconductors.  This includes spin-flip scattering
\cite{bergeret2005-ots,morten2005-stm,morten2004-std} due to magnetic
impurities
$\check{\sigma}_{\rm sf}=-\frac{i}{8\tau_{\rm
    sf}}\vec{S}\tau_3\cdot\check{g}\vec{S}\tau_3$,
spin-orbit scattering
$\check{\sigma}_{\rm so}=-\frac{i}{8\tau_{\rm
    so}}\vec{S}\cdot\check{g}\vec{S}$,
and phonon scattering $\check{\sigma}_{\rm ph}$ (see
Appendix~\ref{sec:eph}).  Below, we parameterize
$\tau_{\rm sf/so}^{-1} = \frac{1 \pm \beta}{2}\tau_{sn}^{-1}$, where
the parameter $-1\le\beta\le1$ describes which of the spin-flip and
spin-orbit scattering mechanisms is stronger.  
For example, the scattering
rates in Al wires were found to be $\tau_{\rm sn}\approx\unit[100]{ps}$ and $\beta\approx0.5$
in Ref.~\onlinecite{poli2008-sir}, so that $T_c\tau_{\rm sn}\sim20$.
We also include orbital dephasing,
$\check{\sigma}_{\rm orb}=-\frac{i}{2\tau_{\rm
    orb}}\tau_3\check{g}\tau_3$,
which is relevant if an external magnetic field is used to generate
the exchange field $h=g\mu_BB$. 
The scattering rate associated with the orbital effect is \cite{belzig1996-ldo}
$\tau_{\rm orb}^{-1}=\frac{T_{c0}\alpha_{\rm orb}}{2}(h/T_{c0})^2$, where
$T_{c0}\approx0.567\Delta_0$ is the BCS critical temperature, and
the parameter $\alpha_{\rm orb}=T_{c0}DW^2/(12g^2\mu_B^2)$ depends on 
the film thickness $W$.

The Keldysh component of $\check g$ can be expressed in terms of the
retarded and advanced matrices and distribution function matrix $\hat{f}$, $\hat{g}^K = \hat{g}^R \hat{f} - \hat{f}\hat{g}^A$. In particular $\hat{f}$  parameterizes the quasiparticle nonequilibrium modes. 
Below, we choose the $z$-axis parallel to the Zeeman field.
In this case, the retarded function is spin-diagonal and we
write it in the form
$\hat{g}^R = \sum_{\sigma=\uparrow/\downarrow}s_{\sigma}[g_{\sigma,1}\tau_1+g_{\sigma,3}\tau_3]$,
where $s_{\uparrow/\downarrow}=[1\pm{}s_z]/2$.
Similarly, we write the distribution function as
$\hat{f} = \sum_{\sigma=\uparrow/\downarrow}s_{\sigma}[f_{L\sigma} + \hat{\tau}_3f_{T\sigma}]$.
The distribution functions $f_{T\sigma}$ characterize the charge imbalance
and $f_{L\sigma}$ the energy imbalance in the two spin bands. An
alternative representation
$f_{T/L}=(f_{T/L,\uparrow}+f_{T/L,\downarrow})/2$,
$f_{L3/T3}=(f_{T/L,\uparrow}-f_{T/L,\downarrow})/2$ was used in
Ref.~\onlinecite{silaev2015-lrs}.

In terms of these functions, Eq.~\eqref{eq:usadel} results in kinetic
equations for the components of $\hat{f}$.  In the steady state they
are rate equations expressing a balance of excitation and relaxation
processes:
\begin{align}
  \label{eq:kinetic-eq}
  \hat{\cal I}_{A}[\hat{f}] 
  + \hat{\cal I}_{\rm sf+so}[\hat{f}] 
  + \hat{\cal I}_{\Delta}[\hat{f}]
  + \hat{\cal I}_{\rm relax}[\hat{f}] 
  = 0
  \,,
\end{align}
where the collision integrals $\hat{\cal I}$ are related to the
corresponding self-energies and $\hat{\Delta}$ via
$\hat{\cal I} = (\hat{g}^R \hat{Z} - \hat{Z}\hat{g}^A)/8$,
$\hat{Z}=i\hat{\sigma}^R\hat{f} - i\hat{f}\hat{\sigma}^A -
i\hat{\sigma}^K$.
Below, we find that $f_{T\sigma}=0$ for our problem, so that
${\cal I}_\Delta=0$.

For the electromagnetic collision integral we get
 \begin{align}
  \hat{Z}_A
  &=
  \frac{DA_0^2}{4}
  \sum_\pm
  \tau_3[g^R_\pm(\hat{f}-\hat{f}_\pm) - (\hat{f}-\hat{f}_\pm)g^A_\pm]\tau_3
  \,,
\end{align}
where $\hat{f}_\pm(E) = \hat{f}(E\pm\omega_0)$.  The
$s_{\uparrow/\downarrow}$ components read
\begin{align}
\label{IA}
  {\cal I}_{A,\sigma}(E)
  &=
  \Tr{}s_\sigma\hat{\cal I}
  =
  \frac{DA_0^2}{4}
  \sum_\pm
  R_{L,\sigma}(E, E \pm \omega)
  \\\notag
  &\qquad\times
  [f_{L,\sigma}(E) - f_{L,\sigma}(E\pm\omega)]
  \,.
\end{align}
The $\hat{\tau}_3s_\sigma$ components vanish, reflecting charge conservation.
The $\pm\omega$ terms indicate driven quasiparticle transitions up/down in energy. 
In terms of the Fermi distribution function $f_\sigma=\frac{1-f_{L\sigma}}{2}$, 
the second line acquires the typical structure for fermion transitions,
 $-2\{f_\sigma(E\pm\omega)[1-f_{\sigma}(E)]-f_\sigma(E)[1-f_{\sigma}(E\pm\omega)]\}$.
The kernel $R$ is
\begin{align}
  R_{L,\sigma}(E, E')
  =
  N_\sigma(E)N_\sigma(E')
  +
  \Im g_{\sigma,1,E}\Im g_{\sigma,1,E'}
  \,,
\end{align}
where $N_\sigma(E)=\Re g_{\sigma,3,E}$ is the spin-dependent density
of states.  The result Eq. (\ref{IA}) is equivalent to a standard photoabsorption
collision integral for each spin.

For the elastic spin-flip and spin-orbit scattering, we have
\begin{align}
  {\cal I}_{{\rm sn},\sigma} 
  &= 
  \frac{S_{\uparrow\downarrow}}{4\tau_{\rm sn}}(f_{L,\sigma}-f_{L,-\sigma})
  \,,
  \\
  S_{\uparrow\downarrow} &= N_{\uparrow}N_{\downarrow} + \beta\Im g_{\uparrow,1}\Im g_{\downarrow,1}
  \,.
\end{align}
To find analytical results we describe inelastic relaxation
within a relaxation-time approximation, for which we
have
\begin{align}
  {\cal I}_{\mathrm{in},\sigma} = \frac{N_\sigma}{2\tau_{\mathrm{in}}}(f_{L\sigma}-f_L^{(0)})
 \,.
\end{align}
We also obtain numerical results with a more detailed model for
electron-phonon scattering (see Appendix~\ref{sec:eph}).

Based on the above equations, we can first solve the components of
$\hat{g}^R$ from Eq.~\eqref{eq:usadel}, which can be done analytically
in some cases, or in general numerically. This provides the coefficients in the
kinetic equations~\eqref{eq:kinetic-eq}.  Alternatively, we also solve
Eq.~\eqref{eq:usadel} directly numerically (see
Appendix~\ref{sec:numerics}), which ensures self-consistency of the
spectral functions.

\section{Spin imbalance}
\label{sec:probes}

\begin{figure}
  \includegraphics{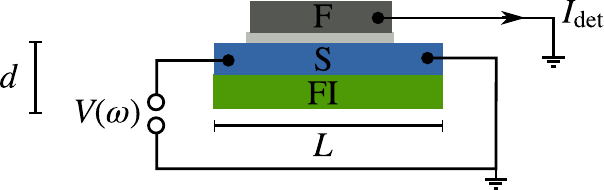}
  \caption{\label{fig:mwcircuit2} 
    Schematic cross-section of a superconductor
    (S) / ferromagnetic insulator (FI) hybrid thin-film structure of
    thickness $d\ll{}L$, driven by an oscillating ac electric field
    corresponding to voltage $V$ at frequency $\omega$.
    The S/FI layer is coupled to a ferromagnetic detector probe (F) 
    via a tunnel junction, where the dc current $I_{\rm det}$ 
    is measured.
  }
\end{figure}

Experimentally, spin imbalance in the superconductor can be probed by electrical
measurements that use spin-filtering probe junctions, for example
ferromagnets. The dc current-voltage relation will in these cases
contain a component that depends on the polarization of the probe and
the spin imbalance in the sample.

The current measured by a spin-filtering tunnel probe
[see Fig.~\ref{fig:mwcircuit2}] coupled to the superconductor and biased
at $V=0$ is given by\cite{bergeret2012-ett}
\begin{align}
  \label{eq:Idet}
  R_{det}I_{det}
  &=
  \mu
  +
  P_{det}\mu_z
  \\
  \mu &=
  \frac{1}{4}\int_{-\infty}^\infty\dd{\epsilon}[N_\uparrow f_{T\uparrow} + N_\downarrow f_{T\downarrow}]
  \,,
  \\
  \mu_z &=
  \frac{1}{4}\int_{-\infty}^\infty\dd{\epsilon}\{
  N_\uparrow [f_{L\uparrow} - f_{L,d}]
  -
  N_\downarrow [f_{L\downarrow} - f_{L,d}]
  \}
  \,,
\end{align}
where $P_{det}$ is the detector polarization in $z$-direction,
$R_{det}$ the junction resistance, and
$f_{L,d}=\tanh\frac{E}{2T_{\rm det}}$ the equilibrium distribution
in the detector.  The current in the detector is a measure of the charge ($\mu$) and spin ($\mu_z$) imbalances.

The ac drive only excites the modes $f_{L,\uparrow/\downarrow}$ which
carry no charge imbalance, so that on this level of analysis, no
photoelectric effect is present ($\mu=0$).  Moreover,
Eq.~\eqref{eq:kinetic-eq} together with a spin-independent relaxation
time yields no spin imbalance ($\mu_z=0$). This follows
from $R_{L,\sigma}(E,E')=R_{L,\sigma}(E',E)$ and
$\mu_z\propto{}\int_{-\infty}^\infty\dd{E}\sum_\sigma\sigma{\cal
  I}_{\mathrm{in},\sigma}\propto\int_{-\infty}^\infty\dd{E}\sum_\sigma\sigma{\cal
  I}_{A,\sigma}(E)=0$, reflecting conservation of spin.

In practice, however, elastic spin-flip scattering cannot be ignored,
and the associated scattering times can be short compared to the
inelastic collisions, $\tau_{\rm sn}\ll{}\tau_{\mathrm{in}}$.  Under
such conditions, ac drive can result to nonzero spin imbalance
$\mu_z\ne0$ in the steady state. Away from the strict diffusive limit,
magnetic impurities result to a photoelectric effect of order
$\ell_{\rm el}^2A_0^2$ also in the absence of the Zeeman
splitting. \cite{zaitsev1986-pes,kalenkov2015-dsb} Here, we 
concentrate only on the diffusive limit and hence the Zeeman splitting is crucial. 

Writing the solution of the kinetic equations
Eqs.~\eqref{eq:kinetic-eq},\eqref{eq:Idet} using the relaxation time
approximation for inelastic processes, and considering the limit of
weak spin-flip scattering
$\tau_{\mathrm{in}}\ll\tau_{sn}$, $DA_0^2\tau_{\mathrm{in}}\ll1$, we find
\begin{align}
  f_{L\sigma}-f_{L\sigma}^{(0)}
  \simeq
  -
  \frac{2{\cal I}_{A,\sigma}^{(0)}\tau_{\mathrm{in}}}{N_\sigma}
  -
  \sigma \frac{S_{\uparrow\downarrow}\tau_{\mathrm{in}}^2}{2\tau_{sn}}
  \frac{N_\uparrow{\cal I}_{A\downarrow}^{(0)}-N_\downarrow{\cal I}_{A\uparrow}^{(0)}}{N_\uparrow{}N_\downarrow}
  \,.
\end{align}
where ${\cal I}_{A\sigma}^{(0)}={\cal I}_{A\sigma}[f_L^{(0)}]$, and
$f_L^{(0)}=\tanh\frac{E}{2T}$ is the equilibrium distribution.  As
noted above, the first term does not contribute to spin imbalance, but
the second term does.  The result is illustrated in
Fig.~\ref{fig:schematic}(b): the transitions driven by the ac field
generate an imbalance of quasiparticles inside both spin bands, which
the spin-flip scattering converts to a spin imbalance at energies
$|E|>|\Delta|+|h|$.

\begin{figure}
  \includegraphics{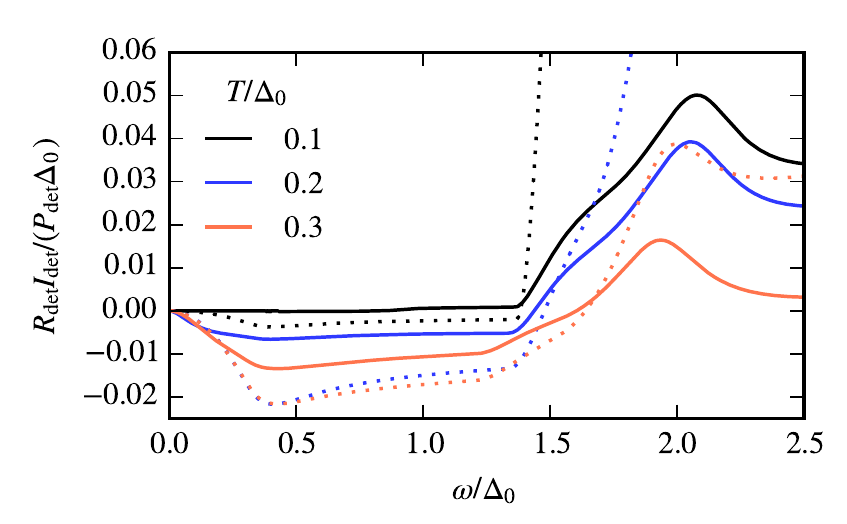}
  \caption{
    \label{fig:Idet}
    Current at the detector probe under ac excitation,
    for $\tau_{sn}T_{c0}=12.5$, $\beta=0.5$, $h/\Delta_0=0.17$, $DA_0^2/\Delta_0=4\times10^{-4}$,
    $\alpha_{\rm orb}=0.1$.
    Solid lines correspond to a numerical solution of Eq.~\eqref{eq:usadel} with
    the phonon model with $\tau_{eph,0}T_{c0}=100$,
    and dotted lines to Eq.~\eqref{eq:Idetapprox} with $\tau_{\mathrm{in}}=(T_{c0}/T)^3\tau_{eph,0}$.
  }
\end{figure}

In a typical situation,  however,  we expect that spin-flip scattering is fast
compared to inelastic relaxation ($\tau_{sn}\ll\tau_{\mathrm{in}}$). In this limit
we have for the detector current,
\begin{align}
  \label{eq:Idetapprox}
  I_{\rm det}
  &=
  -\frac{P_{\rm det}DA_0^2\tau_{\mathrm{in}}}{4R_{\rm det}}
  \int_{-\infty}^\infty\dd{E}
  \sum_\pm
  [f_{L}^{(0)}(E) - f_{L}^{(0)}(E\pm\omega)]
  \\\notag
  &\qquad\times
  \frac{
    N_{\uparrow}(E)R_{L\downarrow}(E,E\pm\omega) - N_\downarrow(E) R_{L\uparrow}(E,E\pm\omega)
  }{
    N_\uparrow(E)+N_\downarrow(E)
  }
  .
\end{align}
The result is shown in Fig.~\ref{fig:Idet} for representative
parameters. The figure also shows results computed numerically using a
phonon model (see Appendix~\ref{sec:eph}). The two are qualitatively
similar, up to differences largely originating from temperature and
energy dependence in the relaxation rates.  As the frequency increases,
the amplitude of the signal also increases up to the point
$\omega\approx{}2h$, where the process depicted in
Fig.~\ref{fig:schematic}(b) saturates.  At large frequencies
$\omega\gtrsim{}2\Delta$ the detector current changes sign, as
microwave-driven pair breaking starts to contribute. In this regime,
the resulting signal can be understood as a thermoelectric current
\cite{machon2013-nte,ozaeta2014-hte,machon2014-gte} driven by a
temperature difference $T_S>T_F$ caused by the  heating of the
superconductor by the drive. 

In addition to the nonequilibrium-generated signal, oscillating
electric fields can introduce a voltage drop $V_{ac}$ across the
detector tunnel barrier.  
This results to an additional signal from photoassisted tunnelling.
Within the above approach, the total current is given by the Tien-Gordon
result, \cite{tien1963-mpo,tucker1985-qda}
\begin{align}
  I = \sum_{n=-\infty}^\infty J_n^2\left(\frac{eV_{ac}}{\hbar\omega_{ac}}\right) 
  I_{\rm det}(V_{dc} + n\hbar\omega_{ac}/e)
  \,,
\end{align}
where $I_{\rm det}(V)$ is the current-voltage relation in the absence of
$V_{ac}$, and $J_n$ are Bessel functions. In the leading order in
driving amplitude ($V_{\rm ac}\to0$) and without dc bias ($V=0$),
\begin{align}
  I
  &\simeq
  I_{\rm det}(0) 
  +
  I_{0,T}
  \\
  I_{0,T} &= \frac{V_{\rm ac}^2}{4\omega^2}[I_{\rm eq}(\omega) + I_{\rm eq}(-\omega)]
  \,,
\end{align}
where $I_{\rm det}(0)$ is the nonequilibrium current~\eqref{eq:Idet},
$I_{0,T}$ the photoassisted current, and $I_{\rm eq}(V)$ is the
IV-relation of the FIS junction when the superconductor is at
equilibrium.  We have
\begin{align}
  I_{0,T}
  &=
  \frac{P_{\rm det}V_{\rm ac}^2}{2R_{\rm det}\omega^2}
  \int_{-\infty}^\infty\dd{E}
  N_0(E)
  \Bigl[
  \\\notag
  &
  f_L^{(0)}(E+h) - \frac{f_L^{(0)}(E-\omega+h) + f_L^{(0)}(E+\omega+h)}{2}
  \\\notag
  &-
  f_L^{(0)}(E-h) + \frac{f_L^{(0)}(E-\omega-h) + f_L^{(0)}(E+\omega-h)}{2}
  \Bigr]
  \,,
\end{align}
where $N_0=\Re\frac{|E|}{\sqrt{E^2-\Delta^2}}$ is the BCS density of
states.  Consider now $\omega,h\lesssim{}T$. We find
\begin{align}
  I_{0,T}
  &=
  -\frac{P_{\rm det}V_{ac}^2h}{2R_{\rm det}\Delta_0^2} \eta(T/\Delta)
  \,,
\end{align}
where $\eta\sim{}1$. The photoassisted tunneling current has the same sign
$I<0$ as a thermoelectric current generated by heating the
ferromagnet, $T_S<T_F$, and therefore also the same sign as the
nonequilibrium effect~\eqref{eq:Idetapprox} at the low
frequencies. The ratio of the photoassisted tunneling to the
nonequilibrium one is
$I_{\rm det}(0)/I_{0,T}\propto(A_0\omega_0\ell_{\mathrm{in}}/V_{ac})^2$ where
$\ell_{\mathrm{in}}=\sqrt{D\tau_{\mathrm{in}}}$.

In principle, $V_{ac}$ can be suppressed by suitable circuit design.
An approach for suppressing the photoassisted tunnelling used in a
previous experiment \cite{horstman1981-gen} measuring the gap
enhancement due to microwave drive was to use a large-area lateral
tunnel junction with a high capacitance.  Assuming the microwave
currents in the S-film and through the detector junction are of
similar magnitude (cf. Ref.~\onlinecite{mooij1983-nem}), one has
$A_0\omega_0\ell_{\mathrm{in}}/V_{ac}\propto{}Z_{S,\ell_{\mathrm{in}}}(\omega)/Z_T(\omega)$,
where $Z_{S,\ell_{\mathrm{in}}}$ is the impedance of the $S$ film of length
$\ell_{\mathrm{in}}$ and $Z_T$ is the junction impedance.  Moreover, the presence
of photoassisted tunneling can in principle be recognized from the
appearance of frequency replicas in the tunneling I-V relation, which
should  not be  present in the nonequilibrium signal.

\section{Gap instability}
\label{sec:gapeqn}

The nonequilibrium spin accumulation affects the magnitude of the
superconducting order parameter, potentially leading to large changes
and collapse of superconductivity for large driving amplitudes.

Let us consider the effect of the driving on the
superconducting order parameter,
\begin{align}
  \label{eq:gapeqn}
  \Delta
  =
  \frac{\lambda}{8}
  \int_{-E_0}^{E_0}\dd{E}
  \tr\frac{\tau_1+i\tau_2}{2}\hat{g}^K(E)
  \,.
\end{align}
Here, we assume singlet pairing, and $\lambda$ is the corresponding
coupling constant and $E_0$ the BCS cutoff.  The simplest situation is
obtained by neglecting spin-flip and spin-orbit scattering. In this
case, we can observe that the only spin structure in the equations
arises from the Zeeman term, $\vec{h}\cdot\vec{S}$. Treating inelastic
collisions within a relaxation time approximation, we find in leading
order
\begin{align}
  \frac{N_\sigma\delta f_{L\sigma}}{\tau_{\mathrm{in}}}
  &=
  -
  \frac{DA_0^2}{2}\sum_{\pm}R_{L\sigma}(E,E\pm\omega) [f_L^{(0)}(E) - f_L^{(0)}(E \pm \omega)]
  \,.
\end{align}
The nonequilibrium part of the gap equation now reads
\begin{align}
  \label{eq:gapeqn-nosf}
  \delta \Delta
  &=
  \frac{\lambda}{4}
  DA_0^2\tau_{\mathrm{in}}
  \int_{-\infty}^\infty dE
  \sum_{\sigma=\pm}\sum_{\pm}\frac{R_{\pm}^{(0)}(E)F_0(E)}{N_0(E)} 
  \\\notag&\qquad
  \times[f_0(E+\sigma h) - f_0(E+\sigma h \pm \omega)]
  \\
  &\simeq
  \frac{\lambda}{4}
  \frac{DA_0^2\tau_{\mathrm{in}}\omega}{2T}
  \int_{-\infty}^\infty dE
  \sum_{\pm}\frac{\pm R_{\pm}^{(0)}(E)F_0(E)}{N_0(E)}
\end{align}
where on the second line we expand around $T\gg{}h,\Delta$, and
$F_0(E) = \frac{1}{4}\Im \tr \tau_1 \hat{g}^R(E)\rvert_{h=0}$. The result is the
same as for zero Zeeman field.  Without spin-flip scattering, the
exchange field does not have a significant effect at high temperatures,
and the result coincides with known results in
Ref.~\onlinecite{eliashberg1970-fss}: the superconducting gap is
enhanced by the driving, and the superconducting branch extends to
$T>T_c$. Numerical calculations also indicate that the exchange field
does not cause significant qualitative changes at lower temperatures
either (see below).

The spin imbalance generated by the spin-flip scattering
however modifies the above conclusion, provided these processes are
not slow compared to energy relaxation. As above, let us now assume
$\tau_{\rm sn}\ll{}\tau_{\mathrm{in}}$. In this case we find
\begin{align}
  \delta\Delta
  &=
  \frac{\lambda}{4} 
  DA_0^2\tau_{\mathrm{in}}
  \int_{-\infty}^\infty\dd{\epsilon}
  \frac{
    F_{\uparrow}+F_{\downarrow}
  }{
    N_{\uparrow}+N_{\downarrow}
  }
  \\\notag
  &\quad\times
  \sum_\pm
  R_{L}(E,E\pm\omega)[f_L^{(0)}(E) - f_{L}^{(0)}(E\pm\omega)]
  \,,
\end{align}
where $R_{L}=\frac{1}{2}\sum_\sigma{}R_{L\sigma}$, and
$F_\sigma=\Re{}g_{\sigma,1}$ is the coherence function.  The
difference to Eq.~\eqref{eq:gapeqn-nosf} is in that the elastic
spin-flip scattering forces the quasiparticle distributions for both
spins to be the same, rather than being copies of a single
distribution shifted by the exchange field.

\begin{figure}
  \includegraphics{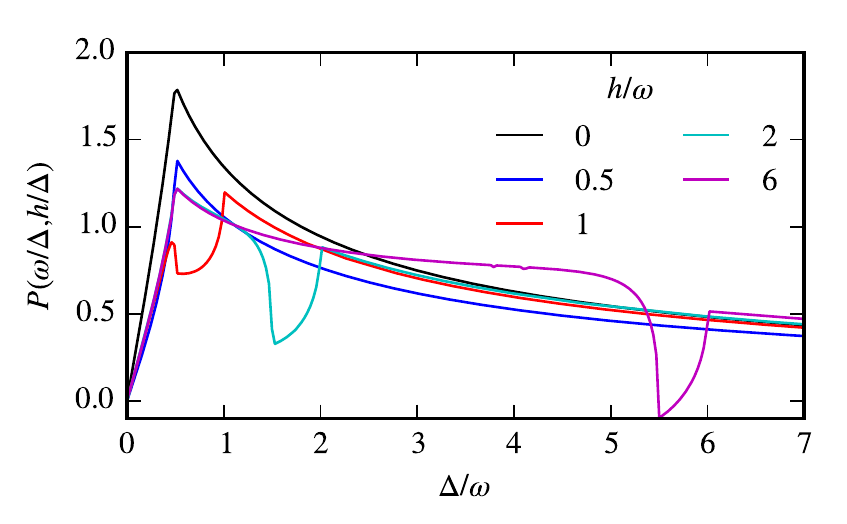}
  \caption{
    \label{fig:P-enhancement}
    Function $P$ for different exchange fields $h$. 
    The resonant dips are located at $\Delta = h$.
  }
\end{figure}

In order to obtain analytical results, let us consider a situation in
which the effect of scattering on the spectral functions is small,
$1/\Delta\ll{}\tau_{\rm sf},\tau_{\rm so},\tau_{\rm orb}$. Then,
$N_\sigma(E)=N(E-\sigma h)$, $F_\sigma(E)=F(E-\sigma h)$,
where $N(E) = N(-E) = \Re[E/\sqrt{E^2-\Delta^2}]$ is the BCS density
of states, and $F(E) = -F(-E) = \Re[\Delta / \sqrt{E^2 - \Delta^2}]$
the BCS coherence function.  Close to the critical temperature
$T\approx{}T_c(h)$ and neglecting the orbital effect, the gap equation
can be expanded to the form \cite{maki1964-pps1,maki1964-pps,kopnin2001-ton}
\begin{align}
  \label{eq:gl}
  \ln\frac{T_{c}}{T}
  =
  \frac{7\zeta(3) - 186\zeta(5)\frac{h^2}{4\pi^2T^2}}{8\pi^2T^2}\Delta^2
  -
  \frac{DA_0^2\tau_{\mathrm{in}}\omega}{4T} P(\frac{\omega}{\Delta}, \frac{h}{\Delta})
  \,.
\end{align}
The nonequlibrium part $\delta\Delta$ results to an extra term,
\cite{eliashberg1970-fss,kopnin2001-ton} whose parameter dependence is
given by the dimensionless function
\begin{align}
  P(w,y)
  &=
  \int_{1}^\infty\dd{x}
  \frac{
    [Z(x,y) - Z(x+w,y)]
    [x(x+w) + 1]
  }{
    \sqrt{x^2 - 1} \sqrt{(x+w)^2 - 1}
  }
  \\\notag
  &+
  \theta(w-2)
  \int_1^{w-1}\dd{x}
  \frac{
    Z(x,y)
    [x(x-w) + 1]
  }{
    \sqrt{x^2 - 1} \sqrt{(x-w)^2 - 1}
  }.
\end{align}
Here,
\begin{align}
  Z(x,y) 
  &= 
  \frac{1}{2}
  \sum_{\alpha=\pm} \frac{
    \sum_{\gamma=\pm}F[x\Delta + (\alpha + \gamma) y\Delta]
  }{
    \sum_{\gamma=\pm}N[x\Delta + (\alpha + \gamma) y\Delta]
  }
  \,.
\end{align}
In the absence of spin splitting, $Z(x,y=0)=1/x$.
The function $P$ is plotted in Fig.~\ref{fig:P-enhancement}.
We can also find its asymptotic behavior for $w\to0$,
\begin{align}
  \label{eq:P-lowfreq}
  P(w,y)
  \simeq
  \begin{cases}
  \sqrt{\frac{2yw}{1+y}}
  +
  \frac{w\ln\frac{8}{w}}{2 + 2y}
  \,,
  & w\ll{}y<1\,,
  \\
  \sqrt{8w} - (\ln\frac{8}{w} - 1)w - \frac{w^{3/2}}{\sqrt{2}}
  \,,
  & w\ll{}1\ll{}y\,,
  \\
  (\ln(8/w) - 1)w,
  &
  y\ll{}w\ll{}1
  \,.
  \end{cases}
\end{align}
For $w\gg1$, on the other hand,
\begin{align}
  \label{eq:P-highfreq}
  P(w,y)
  \simeq
  \begin{cases}
    \pi/w\,,\qquad y\ll{}1\ll{}w
    \,,
    \\
    2/w\,,\qquad{} 1\ll{}w,y ; \; |y-w|\gg{}1
    \,.
  \end{cases}
\end{align}
These limits do not include the feature at $\Delta=h$.

As follows from the gap equation~\eqref{eq:gl}, close to $T=T_c$ the
$\Delta(T)$ relation is determined by
$T/T_c-1 \propto{} P(\Delta/\omega,h/\omega)$.  The order parameter
$\Delta(T)$ is then given by the curves in
Fig.~\ref{fig:P-enhancement}, with $y$-axis $\propto{}T-T_c$.
Additional features in the $\Delta(T)$ relation may appear
at $\Delta(T)=h$.  In particular, the relation is multivalued around
this point, indicating that we can expect discontinuous transitions as
a function of temperature. This is related to changes in possible
relaxation channels around $\Delta\sim{}h$: for $\Delta(T)>h$ the
spin-averaged density of states is gapped at $E<|\Delta|-|h|$, but for
$\Delta(T)<h$ the averaged DOS is gapless, as the spin splitting is
large enough to separate the energy gaps of the two spin species.  The
gap enhancement by microwave driving however continues to increase in the $\Delta(T)<h$
regime as long as $\omega<2\Delta(T)$, and is larger at some
frequencies than without spin splitting.

\begin{figure}
  \includegraphics{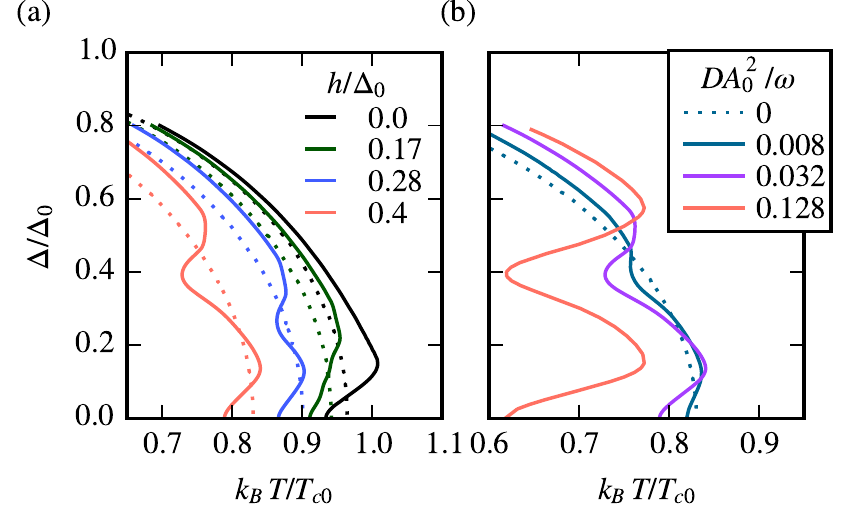}
  \caption{
    \label{fig:gap-vs-A0-h}
    Order parameter $\Delta$ for 
    (a) different exchange fields for $DA_0^2/\omega=0.032$, and
    (b) different drive amplitudes for $h/\Delta_0=0.4$. 
    Dotted lines correspond $DA_0^2=0$.
    Spin-flip scattering time is $\tau_{sn}T_{c0}=12.5$,
    $\beta=0.5$ and $\alpha_{\rm orb}=0.01$,
    $\omega/\Delta_0=0.2$. Inelastic relaxation is described by 
    a phonon model with $\tau_{\text{e-ph},0}T_{c0}=100$.
  }
\end{figure}
  
\begin{figure}
  \includegraphics{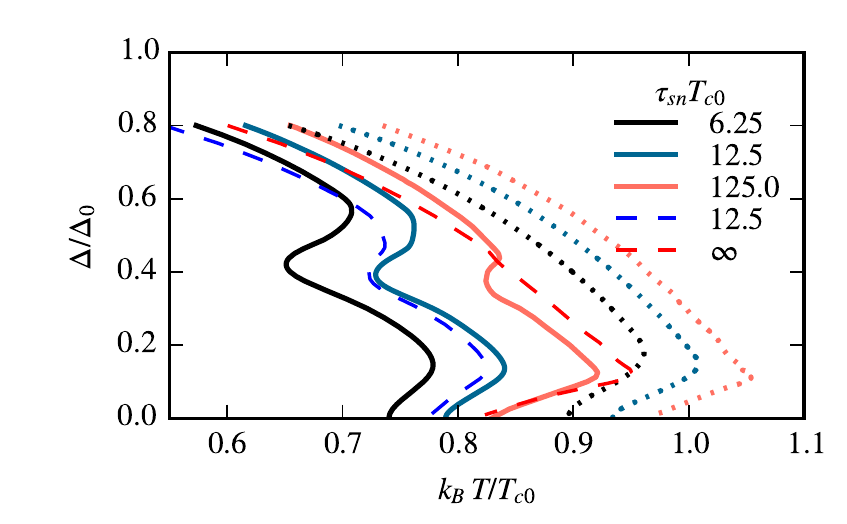}
  \caption{
    \label{fig:gap-vs-tausn}
    Order parameter $\Delta$ for different spin-flip scattering times
    $\tau_{sn}$. Here, $h/\Delta_0=0.4$, and other parameters are as in 
    Fig.~\ref{fig:gap-vs-A0-h}(a).
    The inelastic relaxation is modeled either via phonon model (solid lines)
    or relaxation time approximation (dashed). Results for $h=0$ are also 
    shown (dotted).
  }
\end{figure}

To obtain a more accurate picture, we can solve 
Eqs.~\eqref{eq:usadel} and \eqref{eq:gapeqn}
numerically.
Such results are shown in Figs.~\ref{fig:gap-vs-A0-h}
and~\ref{fig:gap-vs-tausn}.  Figure \ref{fig:gap-vs-A0-h}(a) indicates
the appearance of a multivalued gap, when $h\gtrsim\omega$ and
$\Delta\approx{}h$, as found analytically.  From
Fig.~\ref{fig:gap-vs-tausn} we can note that the feature occurs for a
wide range of scattering times. The feature is absent if there is no
exchange field, and also if there is no spin-flip scattering, so that
the qualitative conclusions based on Eq.~\eqref{eq:gapeqn-nosf} also
apply at lower temperatures.

\section{Discussion}
\label{sec:discussion}

Microwave electric field in a spin-split superconductor drives the
quasiparticles lying above the superconducting gap into a
nonequilibrium state.  In the absence of spin-flipping processes, it
however cannot generate charge or spin imbalance in the diffusive
limit.  Presence of spin-flip scattering enables generation of spin
imbalance, and results to a photoelectric signal observable with
ferromagnetic probes. The effect is closely related to the
thermoelectric effects in magnetic superconductors.  In addition, the
excitation causes an instability in the superconducting order
parameter when the energy gap becomes comparable to the exchange
splitting of the spectrum.

To conclude, we describe production of spin imbalance and a
photo-spin-electric effect in spin-split superconductors, and present
calculations in relevant parameter ranges.  The effects are
experimentally accessible with  state-of-the art methods. A
number of recent experiments in similar systems exist,
\cite{quay2014-fdm,quay2015-qsr,huebler2012-lrs} including also
microwave excitation in the GHz frequency range. \cite{quay2015-qsr}
In addition to using ferromagnetic probes for detecting the photo-spin
signal, the effect can also be seen indirectly by
observing the discontinuous transitions in the superconducting order
parameter. These are experimentally accessible via measurements of the
tunnelling DOS in the superconductor, or for example via measurement
of the supercurrent.

P.V. and T.T.H acknowledge the Academy of Finland for financial
support.  T.T.H. acnowledges funding from the European Research
Council (Grant No. 240362-Heattronics).  The work of F.S.B. 
was supported by Spanish Ministerio de Econom\'ia y Competitividad (MINECO)
 through the Project No. FIS2014-55987-P and Grupos Consolidados UPV/EHU 
 del Gobierno Vasco (Grant No. IT-756-13).

\appendix

\section{Electron-phonon interaction}
\label{sec:eph}

We use a simplified model for inelastic relaxation due to
electron-phonon interaction, \cite{kopnin2001-ton,eliashberg1972-iec}
\begin{align}
  \hat{\sigma}^R(t)
  &=
  -g_{eph}\Bigl\{
  \tilde{D}^R(t) \hat{g}^K(t) 
  + \tilde{D}^K(t) \hat{g}^R(t)
  \Bigr\}
  \\
  \hat{\sigma}^K(t)
  &=
  -g_{eph}\Bigl\{
  [
  \tilde{D}^R(t) - \tilde{D}^A(t)] [\hat{g}^R(t) - \hat{g}^A(t)]
  \\\notag
  &\qquad\qquad
  + \tilde{D}^K(t) \hat{g}^K(t)
  \Bigr\}
  \,,
\end{align}
where the Fourier transformed functions
\begin{align}
  \tilde{D}^{R/A}(\omega) &= \pm i\omega^2 \sgn\omega
  \,,
  \\
  \tilde{D}^{K}(\omega) &= [\tilde{D}^{R}(\omega) - \tilde{D}^{A}(\omega)]\coth\frac{\omega}{2T_{\rm ph}}
  \,,
\end{align}
arise from weighed Fermi surface averages of the phonon Green
functions. Parts that do not contribute to the collision integral have
been subtracted. The collision integral assumes a standard form,
\begin{align}
  {\cal I}_{L\sigma}
  &=
  g_{eph}
  \int_{-\infty}^\infty\frac{\dd{\omega}}{8\pi}
  \omega|\omega|\tr[\hat{g}^{RA}(E)\hat{g}^{RA}(E-\omega)s_\sigma]
  \\\notag&\times
  \{
  f_{L\sigma,E}f_{L\sigma,E-\omega} - 1 + [f_{L\sigma,E}-f_{L\sigma,E-\omega}]\coth\frac{\omega}{2T_{\rm ph}}
  \}
\end{align}
where $\hat{g}^{RA}=\hat{g}^R - \hat{g}^A$.  The prefactor $g_{eph}$
can be defined in terms of a relaxation rate
\begin{align}
  \tau_{eph,0}^{-1} = 4\frac{7\zeta(3)g_{eph}}{\pi} T_{c,0}^3
  \,,
\end{align}
at temperature $T_{c,0}$ in the normal state at
Fermi surface, where $\hat{\sigma}^R=-i\hat{\tau}_3/(2\tau_{\rm eph,0})$.

\section{Numerical details}
\label{sec:numerics}

We solve Eq.~\eqref{eq:usadel} numerically via a Jacobian-free
Newton-GMRES method.  \cite{knoll2003-jnm} To obtain a preconditioner,
we use automatic differentiation to compute the Jacobian of the
energy-local terms, excluding the self-energy parts
$[\check{\sigma}_{\rm eph} + \check{\sigma}_A,\check{g}]$. The energy
convolutions in the self-energies, after energy discretization, are
computed via fast Fourier transforms.

\bibliography{spinacpaper}

\begin{thebibliography}{39}%
\makeatletter
\providecommand \@ifxundefined [1]{%
 \@ifx{#1\undefined}
}%
\providecommand \@ifnum [1]{%
 \ifnum #1\expandafter \@firstoftwo
 \else \expandafter \@secondoftwo
 \fi
}%
\providecommand \@ifx [1]{%
 \ifx #1\expandafter \@firstoftwo
 \else \expandafter \@secondoftwo
 \fi
}%
\providecommand \natexlab [1]{#1}%
\providecommand \enquote  [1]{``#1''}%
\providecommand \bibnamefont  [1]{#1}%
\providecommand \bibfnamefont [1]{#1}%
\providecommand \citenamefont [1]{#1}%
\providecommand \href@noop [0]{\@secondoftwo}%
\providecommand \href [0]{\begingroup \@sanitize@url \@href}%
\providecommand \@href[1]{\@@startlink{#1}\@@href}%
\providecommand \@@href[1]{\endgroup#1\@@endlink}%
\providecommand \@sanitize@url [0]{\catcode `\\12\catcode `\$12\catcode
  `\&12\catcode `\#12\catcode `\^12\catcode `\_12\catcode `\%12\relax}%
\providecommand \@@startlink[1]{}%
\providecommand \@@endlink[0]{}%
\providecommand \url  [0]{\begingroup\@sanitize@url \@url }%
\providecommand \@url [1]{\endgroup\@href {#1}{\urlprefix }}%
\providecommand \urlprefix  [0]{URL }%
\providecommand \Eprint [0]{\href }%
\providecommand \doibase [0]{http://dx.doi.org/}%
\providecommand \selectlanguage [0]{\@gobble}%
\providecommand \bibinfo  [0]{\@secondoftwo}%
\providecommand \bibfield  [0]{\@secondoftwo}%
\providecommand \translation [1]{[#1]}%
\providecommand \BibitemOpen [0]{}%
\providecommand \bibitemStop [0]{}%
\providecommand \bibitemNoStop [0]{.\EOS\space}%
\providecommand \EOS [0]{\spacefactor3000\relax}%
\providecommand \BibitemShut  [1]{\csname bibitem#1\endcsname}%
\let\auto@bib@innerbib\@empty
\bibitem [{\citenamefont {Linder}\ and\ \citenamefont
  {Robinson}(2015)}]{linder2015-ss}%
  \BibitemOpen
  \bibfield  {author} {\bibinfo {author} {\bibfnamefont {J.}~\bibnamefont
  {Linder}}\ and\ \bibinfo {author} {\bibfnamefont {J.~W.~A.}\ \bibnamefont
  {Robinson}},\ }\href {\doibase 10.1038/nphys3242} {\bibfield  {journal}
  {\bibinfo  {journal} {Nature Physics}\ }\textbf {\bibinfo {volume} {11}},\
  \bibinfo {pages} {307} (\bibinfo {year} {2015})}\BibitemShut {NoStop}%
\bibitem [{\citenamefont {Kivelson}\ and\ \citenamefont
  {Rokhsar}(1990)}]{kivelson1990-bqs}%
  \BibitemOpen
  \bibfield  {author} {\bibinfo {author} {\bibfnamefont {S.~A.}\ \bibnamefont
  {Kivelson}}\ and\ \bibinfo {author} {\bibfnamefont {D.~S.}\ \bibnamefont
  {Rokhsar}},\ }\href {\doibase 10.1103/PhysRevB.41.11693} {\bibfield
  {journal} {\bibinfo  {journal} {Phys. Rev. B}\ }\textbf {\bibinfo {volume}
  {41}},\ \bibinfo {pages} {11693} (\bibinfo {year} {1990})}\BibitemShut
  {NoStop}%
\bibitem [{\citenamefont {Zhao}\ and\ \citenamefont
  {Hershfield}(1995)}]{zhao1995-trs}%
  \BibitemOpen
  \bibfield  {author} {\bibinfo {author} {\bibfnamefont {H.~L.}\ \bibnamefont
  {Zhao}}\ and\ \bibinfo {author} {\bibfnamefont {S.}~\bibnamefont
  {Hershfield}},\ }\href {\doibase 10.1103/PhysRevB.52.3632} {\bibfield
  {journal} {\bibinfo  {journal} {Phys. Rev. B}\ }\textbf {\bibinfo {volume}
  {52}},\ \bibinfo {pages} {3632} (\bibinfo {year} {1995})}\BibitemShut
  {NoStop}%
\bibitem [{\citenamefont {Quay}\ \emph {et~al.}(2013)\citenamefont {Quay},
  \citenamefont {Chevallier}, \citenamefont {Bena},\ and\ \citenamefont
  {Aprili}}]{quay2013-sia}%
  \BibitemOpen
  \bibfield  {author} {\bibinfo {author} {\bibfnamefont {C.~H.~L.}\
  \bibnamefont {Quay}}, \bibinfo {author} {\bibfnamefont {D.}~\bibnamefont
  {Chevallier}}, \bibinfo {author} {\bibfnamefont {C.}~\bibnamefont {Bena}}, \
  and\ \bibinfo {author} {\bibfnamefont {M.}~\bibnamefont {Aprili}},\ }\href
  {\doibase 10.1038/nphys2518} {\bibfield  {journal} {\bibinfo  {journal}
  {Nature Physics}\ }\textbf {\bibinfo {volume} {9}},\ \bibinfo {pages} {84}
  (\bibinfo {year} {2013})}\BibitemShut {NoStop}%
\bibitem [{\citenamefont {H\"ubler}\ \emph {et~al.}(2012)\citenamefont
  {H\"ubler}, \citenamefont {Wolf}, \citenamefont {Beckmann},\ and\
  \citenamefont {v.~L\"ohneysen}}]{huebler2012-lrs}%
  \BibitemOpen
  \bibfield  {author} {\bibinfo {author} {\bibfnamefont {F.}~\bibnamefont
  {H\"ubler}}, \bibinfo {author} {\bibfnamefont {M.~J.}\ \bibnamefont {Wolf}},
  \bibinfo {author} {\bibfnamefont {D.}~\bibnamefont {Beckmann}}, \ and\
  \bibinfo {author} {\bibfnamefont {H.}~\bibnamefont {v.~L\"ohneysen}},\ }\href
  {\doibase 10.1103/PhysRevLett.109.207001} {\bibfield  {journal} {\bibinfo
  {journal} {Phys. Rev. Lett.}\ }\textbf {\bibinfo {volume} {109}},\ \bibinfo
  {pages} {207001} (\bibinfo {year} {2012})}\BibitemShut {NoStop}%
\bibitem [{\citenamefont {Wolf}\ \emph {et~al.}(2013)\citenamefont {Wolf},
  \citenamefont {H\"ubler}, \citenamefont {Kolenda}, \citenamefont
  {v.~L\"ohneysen},\ and\ \citenamefont {Beckmann}}]{wolf2013-sin}%
  \BibitemOpen
  \bibfield  {author} {\bibinfo {author} {\bibfnamefont {M.~J.}\ \bibnamefont
  {Wolf}}, \bibinfo {author} {\bibfnamefont {F.}~\bibnamefont {H\"ubler}},
  \bibinfo {author} {\bibfnamefont {S.}~\bibnamefont {Kolenda}}, \bibinfo
  {author} {\bibfnamefont {H.}~\bibnamefont {v.~L\"ohneysen}}, \ and\ \bibinfo
  {author} {\bibfnamefont {D.}~\bibnamefont {Beckmann}},\ }\href {\doibase
  10.1103/PhysRevB.87.024517} {\bibfield  {journal} {\bibinfo  {journal} {Phys.
  Rev. B}\ }\textbf {\bibinfo {volume} {87}},\ \bibinfo {pages} {024517}
  (\bibinfo {year} {2013})}\BibitemShut {NoStop}%
\bibitem [{\citenamefont {Wolf}\ \emph {et~al.}(2014)\citenamefont {Wolf},
  \citenamefont {S\"urgers}, \citenamefont {Fischer},\ and\ \citenamefont
  {Beckmann}}]{wolf2014-spq}%
  \BibitemOpen
  \bibfield  {author} {\bibinfo {author} {\bibfnamefont {M.~J.}\ \bibnamefont
  {Wolf}}, \bibinfo {author} {\bibfnamefont {C.}~\bibnamefont {S\"urgers}},
  \bibinfo {author} {\bibfnamefont {G.}~\bibnamefont {Fischer}}, \ and\
  \bibinfo {author} {\bibfnamefont {D.}~\bibnamefont {Beckmann}},\ }\href
  {\doibase 10.1103/PhysRevB.90.144509} {\bibfield  {journal} {\bibinfo
  {journal} {Phys. Rev. B}\ }\textbf {\bibinfo {volume} {90}},\ \bibinfo
  {pages} {144509} (\bibinfo {year} {2014})}\BibitemShut {NoStop}%
\bibitem [{\citenamefont {Quay}\ \emph {et~al.}(2014)\citenamefont {Quay},
  \citenamefont {Dutreix}, \citenamefont {Chevallier}, \citenamefont {Bena},\
  and\ \citenamefont {Aprili}}]{quay2014-fdm}%
  \BibitemOpen
  \bibfield  {author} {\bibinfo {author} {\bibfnamefont {C.~H.~L.}\
  \bibnamefont {Quay}}, \bibinfo {author} {\bibfnamefont {C.}~\bibnamefont
  {Dutreix}}, \bibinfo {author} {\bibfnamefont {D.}~\bibnamefont {Chevallier}},
  \bibinfo {author} {\bibfnamefont {C.}~\bibnamefont {Bena}}, \ and\ \bibinfo
  {author} {\bibfnamefont {M.}~\bibnamefont {Aprili}},\ }\href@noop {} {}
  (\bibinfo {year} {2014}),\ \bibinfo {note} {arXiv:1408.1832}\BibitemShut
  {NoStop}%
\bibitem [{\citenamefont {Quay}\ \emph {et~al.}(2015)\citenamefont {Quay},
  \citenamefont {Chiffaudel}, \citenamefont {Strunk},\ and\ \citenamefont
  {Aprili}}]{quay2015-qsr}%
  \BibitemOpen
  \bibfield  {author} {\bibinfo {author} {\bibfnamefont {C.~H.~L.}\
  \bibnamefont {Quay}}, \bibinfo {author} {\bibfnamefont {Y.}~\bibnamefont
  {Chiffaudel}}, \bibinfo {author} {\bibfnamefont {C.}~\bibnamefont {Strunk}},
  \ and\ \bibinfo {author} {\bibfnamefont {M.}~\bibnamefont {Aprili}},\
  }\href@noop {} {} (\bibinfo {year} {2015}),\ \bibinfo {note}
  {arXiv:1504.01615}\BibitemShut {NoStop}%
\bibitem [{\citenamefont {Tedrow}\ and\ \citenamefont
  {Meservey}(1971)}]{tedrow1971-sdt}%
  \BibitemOpen
  \bibfield  {author} {\bibinfo {author} {\bibfnamefont {P.~M.}\ \bibnamefont
  {Tedrow}}\ and\ \bibinfo {author} {\bibfnamefont {R.}~\bibnamefont
  {Meservey}},\ }\href {\doibase 10.1103/PhysRevLett.26.192} {\bibfield
  {journal} {\bibinfo  {journal} {Phys. Rev. Lett.}\ }\textbf {\bibinfo
  {volume} {26}},\ \bibinfo {pages} {192} (\bibinfo {year} {1971})}\BibitemShut
  {NoStop}%
\bibitem [{\citenamefont {Tedrow}\ \emph {et~al.}(1986)\citenamefont {Tedrow},
  \citenamefont {Tkaczyk},\ and\ \citenamefont {Kumar}}]{tedrow1986-spe}%
  \BibitemOpen
  \bibfield  {author} {\bibinfo {author} {\bibfnamefont {P.~M.}\ \bibnamefont
  {Tedrow}}, \bibinfo {author} {\bibfnamefont {J.~E.}\ \bibnamefont {Tkaczyk}},
  \ and\ \bibinfo {author} {\bibfnamefont {A.}~\bibnamefont {Kumar}},\ }\href
  {\doibase 10.1103/PhysRevLett.56.1746} {\bibfield  {journal} {\bibinfo
  {journal} {Phys. Rev. Lett.}\ }\textbf {\bibinfo {volume} {56}},\ \bibinfo
  {pages} {1746} (\bibinfo {year} {1986})}\BibitemShut {NoStop}%
\bibitem [{\citenamefont {Silaev}\ \emph {et~al.}(2015)\citenamefont {Silaev},
  \citenamefont {Virtanen}, \citenamefont {Bergeret},\ and\ \citenamefont
  {Heikkil\"a}}]{silaev2015-lrs}%
  \BibitemOpen
  \bibfield  {author} {\bibinfo {author} {\bibfnamefont {M.}~\bibnamefont
  {Silaev}}, \bibinfo {author} {\bibfnamefont {P.}~\bibnamefont {Virtanen}},
  \bibinfo {author} {\bibfnamefont {F.~S.}\ \bibnamefont {Bergeret}}, \ and\
  \bibinfo {author} {\bibfnamefont {T.~T.}\ \bibnamefont {Heikkil\"a}},\ }\href
  {\doibase 10.1103/PhysRevLett.114.167002} {\bibfield  {journal} {\bibinfo
  {journal} {Phys. Rev. Lett.}\ }\textbf {\bibinfo {volume} {114}},\ \bibinfo
  {pages} {167002} (\bibinfo {year} {2015})}\BibitemShut {NoStop}%
\bibitem [{\citenamefont {Bobkova}\ and\ \citenamefont
  {Bobkov}(2015)}]{bobkova2015-lrs}%
  \BibitemOpen
  \bibfield  {author} {\bibinfo {author} {\bibfnamefont {I.~V.}\ \bibnamefont
  {Bobkova}}\ and\ \bibinfo {author} {\bibfnamefont {A.~M.}\ \bibnamefont
  {Bobkov}},\ }\href {\doibase 10.7868/S0370274X15020101} {\bibfield  {journal}
  {\bibinfo  {journal} {Pis'ma Zh. Eksp. Teor. Fiz.}\ }\textbf {\bibinfo
  {volume} {101}},\ \bibinfo {pages} {124} (\bibinfo {year}
  {2015})}\BibitemShut {NoStop}%
\bibitem [{\citenamefont {Kalenkov}\ \emph {et~al.}(2012)\citenamefont
  {Kalenkov}, \citenamefont {Zaikin},\ and\ \citenamefont
  {Kuzmin}}]{kalenkov2012-tlt}%
  \BibitemOpen
  \bibfield  {author} {\bibinfo {author} {\bibfnamefont {M.~S.}\ \bibnamefont
  {Kalenkov}}, \bibinfo {author} {\bibfnamefont {A.~D.}\ \bibnamefont
  {Zaikin}}, \ and\ \bibinfo {author} {\bibfnamefont {L.~S.}\ \bibnamefont
  {Kuzmin}},\ }\href {\doibase 10.1103/PhysRevLett.109.147004} {\bibfield
  {journal} {\bibinfo  {journal} {Phys. Rev. Lett.}\ }\textbf {\bibinfo
  {volume} {109}},\ \bibinfo {pages} {147004} (\bibinfo {year}
  {2012})}\BibitemShut {NoStop}%
\bibitem [{\citenamefont {Machon}\ \emph {et~al.}(2013)\citenamefont {Machon},
  \citenamefont {Eschrig},\ and\ \citenamefont {Belzig}}]{machon2013-nte}%
  \BibitemOpen
  \bibfield  {author} {\bibinfo {author} {\bibfnamefont {P.}~\bibnamefont
  {Machon}}, \bibinfo {author} {\bibfnamefont {M.}~\bibnamefont {Eschrig}}, \
  and\ \bibinfo {author} {\bibfnamefont {W.}~\bibnamefont {Belzig}},\
  }\href@noop {} {\bibfield  {journal} {\bibinfo  {journal} {Phys. Rev. Lett.}\
  }\textbf {\bibinfo {volume} {110}},\ \bibinfo {pages} {047002} (\bibinfo
  {year} {2013})}\BibitemShut {NoStop}%
\bibitem [{\citenamefont {Kalenkov}\ and\ \citenamefont
  {Zaikin}(2014)}]{kalenkov2014-ehi}%
  \BibitemOpen
  \bibfield  {author} {\bibinfo {author} {\bibfnamefont {M.~S.}\ \bibnamefont
  {Kalenkov}}\ and\ \bibinfo {author} {\bibfnamefont {A.~D.}\ \bibnamefont
  {Zaikin}},\ }\href {\doibase 10.1103/PhysRevB.90.134502} {\bibfield
  {journal} {\bibinfo  {journal} {Phys. Rev. B}\ }\textbf {\bibinfo {volume}
  {90}},\ \bibinfo {pages} {134502} (\bibinfo {year} {2014})}\BibitemShut
  {NoStop}%
\bibitem [{\citenamefont {Machon}\ \emph {et~al.}(2014)\citenamefont {Machon},
  \citenamefont {Eschrig},\ and\ \citenamefont {Belzig}}]{machon2014-gte}%
  \BibitemOpen
  \bibfield  {author} {\bibinfo {author} {\bibfnamefont {P.}~\bibnamefont
  {Machon}}, \bibinfo {author} {\bibfnamefont {M.}~\bibnamefont {Eschrig}}, \
  and\ \bibinfo {author} {\bibfnamefont {W.}~\bibnamefont {Belzig}},\ }\href
  {\doibase 10.1088/1367-2630/16/7/073002} {\bibfield  {journal} {\bibinfo
  {journal} {New J. Phys.}\ }\textbf {\bibinfo {volume} {16}},\ \bibinfo
  {pages} {073002} (\bibinfo {year} {2014})}\BibitemShut {NoStop}%
\bibitem [{\citenamefont {Ozaeta}\ \emph {et~al.}(2014)\citenamefont {Ozaeta},
  \citenamefont {Virtanen}, \citenamefont {Bergeret},\ and\ \citenamefont
  {Heikkil\"a}}]{ozaeta2014-hte}%
  \BibitemOpen
  \bibfield  {author} {\bibinfo {author} {\bibfnamefont {A.}~\bibnamefont
  {Ozaeta}}, \bibinfo {author} {\bibfnamefont {P.}~\bibnamefont {Virtanen}},
  \bibinfo {author} {\bibfnamefont {F.~S.}\ \bibnamefont {Bergeret}}, \ and\
  \bibinfo {author} {\bibfnamefont {T.~T.}\ \bibnamefont {Heikkil\"a}},\
  }\href@noop {} {\bibfield  {journal} {\bibinfo  {journal} {Phys. Rev. Lett.}\
  }\textbf {\bibinfo {volume} {112}},\ \bibinfo {pages} {057001} (\bibinfo
  {year} {2014})}\BibitemShut {NoStop}%
\bibitem [{\citenamefont {Kolenda}\ \emph {et~al.}(2015)\citenamefont
  {Kolenda}, \citenamefont {Wolf},\ and\ \citenamefont
  {Beckmann}}]{kolenda2015-otc}%
  \BibitemOpen
  \bibfield  {author} {\bibinfo {author} {\bibfnamefont {S.}~\bibnamefont
  {Kolenda}}, \bibinfo {author} {\bibfnamefont {M.~J.}\ \bibnamefont {Wolf}}, \
  and\ \bibinfo {author} {\bibfnamefont {D.}~\bibnamefont {Beckmann}},\
  }\href@noop {} {} (\bibinfo {year} {2015}),\ \bibinfo {note}
  {arXiv:1509.05568}\BibitemShut {NoStop}%
\bibitem [{\citenamefont {Zaitsev}(1986)}]{zaitsev1986-pes}%
  \BibitemOpen
  \bibfield  {author} {\bibinfo {author} {\bibfnamefont {A.~V.}\ \bibnamefont
  {Zaitsev}},\ }\href@noop {} {\bibfield  {journal} {\bibinfo  {journal} {Sov.
  Phys. JETP}\ }\textbf {\bibinfo {volume} {63}},\ \bibinfo {pages} {579}
  (\bibinfo {year} {1986})}\BibitemShut {NoStop}%
\bibitem [{\citenamefont {Kalenkov}\ and\ \citenamefont
  {Zaikin}(2015)}]{kalenkov2015-dsb}%
  \BibitemOpen
  \bibfield  {author} {\bibinfo {author} {\bibfnamefont {M.~S.}\ \bibnamefont
  {Kalenkov}}\ and\ \bibinfo {author} {\bibfnamefont {A.~D.}\ \bibnamefont
  {Zaikin}},\ }\href {\doibase 10.1103/PhysRevB.92.014507} {\bibfield
  {journal} {\bibinfo  {journal} {Phys. Rev. B}\ }\textbf {\bibinfo {volume}
  {92}},\ \bibinfo {pages} {014507} (\bibinfo {year} {2015})}\BibitemShut
  {NoStop}%
\bibitem [{\citenamefont {Eliashberg}(1970)}]{eliashberg1970-fss}%
  \BibitemOpen
  \bibfield  {author} {\bibinfo {author} {\bibfnamefont {G.~M.}\ \bibnamefont
  {Eliashberg}},\ }\href@noop {} {\bibfield  {journal} {\bibinfo  {journal}
  {JETP Lett.}\ }\textbf {\bibinfo {volume} {11}},\ \bibinfo {pages} {114}
  (\bibinfo {year} {1970})}\BibitemShut {NoStop}%
\bibitem [{\citenamefont {Miao}\ and\ \citenamefont
  {Moodera}(2015)}]{miao2015-smm}%
  \BibitemOpen
  \bibfield  {author} {\bibinfo {author} {\bibfnamefont {G.-X.}\ \bibnamefont
  {Miao}}\ and\ \bibinfo {author} {\bibfnamefont {J.~S.}\ \bibnamefont
  {Moodera}},\ }\href {\doibase 10.1039/C4CP04599H} {\bibfield  {journal}
  {\bibinfo  {journal} {Phys. Chem. Chem. Phys.}\ }\textbf {\bibinfo {volume}
  {17}},\ \bibinfo {pages} {751} (\bibinfo {year} {2015})}\BibitemShut
  {NoStop}%
\bibitem [{\citenamefont {Bergeret}\ \emph {et~al.}(2005)\citenamefont
  {Bergeret}, \citenamefont {Volkov},\ and\ \citenamefont
  {Efetov}}]{bergeret2005-ots}%
  \BibitemOpen
  \bibfield  {author} {\bibinfo {author} {\bibfnamefont {F.~S.}\ \bibnamefont
  {Bergeret}}, \bibinfo {author} {\bibfnamefont {A.~F.}\ \bibnamefont
  {Volkov}}, \ and\ \bibinfo {author} {\bibfnamefont {K.~B.}\ \bibnamefont
  {Efetov}},\ }\href {\doibase 10.1103/RevModPhys.77.1321} {\bibfield
  {journal} {\bibinfo  {journal} {Rev. Mod. Phys.}\ }\textbf {\bibinfo {volume}
  {77}},\ \bibinfo {pages} {1321} (\bibinfo {year} {2005})}\BibitemShut
  {NoStop}%
\bibitem [{\citenamefont {Usadel}(1970)}]{usadel1970-gde}%
  \BibitemOpen
  \bibfield  {author} {\bibinfo {author} {\bibfnamefont {K.~D.}\ \bibnamefont
  {Usadel}},\ }\href@noop {} {\bibfield  {journal} {\bibinfo  {journal}
  {Phys.~Rev.~Lett.}\ }\textbf {\bibinfo {volume} {25}},\ \bibinfo {pages}
  {507} (\bibinfo {year} {1970})}\BibitemShut {NoStop}%
\bibitem [{\citenamefont {Kopnin}(2001)}]{kopnin2001-ton}%
  \BibitemOpen
  \bibfield  {author} {\bibinfo {author} {\bibfnamefont {N.~B.}\ \bibnamefont
  {Kopnin}},\ }\href@noop {} {}\bibinfo {series} {International series of
  monographs on physics}\ No.\ \bibinfo {number} {110}\ (\bibinfo  {publisher}
  {Oxford University Press},\ \bibinfo {year} {2001})\BibitemShut {NoStop}%
\bibitem [{\citenamefont {Morten}\ \emph {et~al.}(2004)\citenamefont {Morten},
  \citenamefont {Brataas},\ and\ \citenamefont {Belzig}}]{morten2004-std}%
  \BibitemOpen
  \bibfield  {author} {\bibinfo {author} {\bibfnamefont {J.~P.}\ \bibnamefont
  {Morten}}, \bibinfo {author} {\bibfnamefont {A.}~\bibnamefont {Brataas}}, \
  and\ \bibinfo {author} {\bibfnamefont {W.}~\bibnamefont {Belzig}},\ }\href
  {\doibase 10.1103/PhysRevB.70.212508} {\bibfield  {journal} {\bibinfo
  {journal} {Phys. Rev. B}\ }\textbf {\bibinfo {volume} {70}},\ \bibinfo
  {pages} {212508} (\bibinfo {year} {2004})}\BibitemShut {NoStop}%
\bibitem [{\citenamefont {Morten}\ \emph {et~al.}(2005)\citenamefont {Morten},
  \citenamefont {Brataas},\ and\ \citenamefont {Belzig}}]{morten2005-stm}%
  \BibitemOpen
  \bibfield  {author} {\bibinfo {author} {\bibfnamefont {J.~P.}\ \bibnamefont
  {Morten}}, \bibinfo {author} {\bibfnamefont {A.}~\bibnamefont {Brataas}}, \
  and\ \bibinfo {author} {\bibfnamefont {W.}~\bibnamefont {Belzig}},\ }\href
  {\doibase 10.1103/PhysRevB.72.014510} {\bibfield  {journal} {\bibinfo
  {journal} {Phys. Rev. B}\ }\textbf {\bibinfo {volume} {72}},\ \bibinfo
  {pages} {014510} (\bibinfo {year} {2005})}\BibitemShut {NoStop}%
\bibitem [{\citenamefont {Poli}\ \emph {et~al.}(2008)\citenamefont {Poli},
  \citenamefont {Morten}, \citenamefont {Urech}, \citenamefont {Brataas},
  \citenamefont {Haviland},\ and\ \citenamefont {Korenivski}}]{poli2008-sir}%
  \BibitemOpen
  \bibfield  {author} {\bibinfo {author} {\bibfnamefont {N.}~\bibnamefont
  {Poli}}, \bibinfo {author} {\bibfnamefont {J.~P.}\ \bibnamefont {Morten}},
  \bibinfo {author} {\bibfnamefont {M.}~\bibnamefont {Urech}}, \bibinfo
  {author} {\bibfnamefont {A.}~\bibnamefont {Brataas}}, \bibinfo {author}
  {\bibfnamefont {D.~B.}\ \bibnamefont {Haviland}}, \ and\ \bibinfo {author}
  {\bibfnamefont {V.}~\bibnamefont {Korenivski}},\ }\href {\doibase
  10.1103/PhysRevLett.100.136601} {\bibfield  {journal} {\bibinfo  {journal}
  {Phys. Rev. Lett.}\ }\textbf {\bibinfo {volume} {100}},\ \bibinfo {pages}
  {136601} (\bibinfo {year} {2008})}\BibitemShut {NoStop}%
\bibitem [{\citenamefont {Belzig}\ \emph {et~al.}(1996)\citenamefont {Belzig},
  \citenamefont {Bruder},\ and\ \citenamefont {Sch\"on}}]{belzig1996-ldo}%
  \BibitemOpen
  \bibfield  {author} {\bibinfo {author} {\bibfnamefont {W.}~\bibnamefont
  {Belzig}}, \bibinfo {author} {\bibfnamefont {C.}~\bibnamefont {Bruder}}, \
  and\ \bibinfo {author} {\bibfnamefont {G.}~\bibnamefont {Sch\"on}},\ }\href
  {\doibase 10.1103/PhysRevB.54.9443} {\bibfield  {journal} {\bibinfo
  {journal} {Phys. Rev. B}\ }\textbf {\bibinfo {volume} {54}},\ \bibinfo
  {pages} {9443} (\bibinfo {year} {1996})}\BibitemShut {NoStop}%
\bibitem [{\citenamefont {Bergeret}\ \emph {et~al.}(2012)\citenamefont
  {Bergeret}, \citenamefont {Verso},\ and\ \citenamefont
  {Volkov}}]{bergeret2012-ett}%
  \BibitemOpen
  \bibfield  {author} {\bibinfo {author} {\bibfnamefont {F.~S.}\ \bibnamefont
  {Bergeret}}, \bibinfo {author} {\bibfnamefont {A.}~\bibnamefont {Verso}}, \
  and\ \bibinfo {author} {\bibfnamefont {A.~F.}\ \bibnamefont {Volkov}},\
  }\href {\doibase 10.1103/PhysRevB.86.214516} {\bibfield  {journal} {\bibinfo
  {journal} {Phys. Rev. B}\ }\textbf {\bibinfo {volume} {86}},\ \bibinfo
  {pages} {214516} (\bibinfo {year} {2012})}\BibitemShut {NoStop}%
\bibitem [{\citenamefont {Tien}\ and\ \citenamefont
  {Gordon}(1963)}]{tien1963-mpo}%
  \BibitemOpen
  \bibfield  {author} {\bibinfo {author} {\bibfnamefont {P.~K.}\ \bibnamefont
  {Tien}}\ and\ \bibinfo {author} {\bibfnamefont {J.~P.}\ \bibnamefont
  {Gordon}},\ }\href {\doibase 10.1103/PhysRev.129.647} {\bibfield  {journal}
  {\bibinfo  {journal} {Phys. Rev.}\ }\textbf {\bibinfo {volume} {129}},\
  \bibinfo {pages} {647} (\bibinfo {year} {1963})}\BibitemShut {NoStop}%
\bibitem [{\citenamefont {Tucker}\ and\ \citenamefont
  {Feldman}(1985)}]{tucker1985-qda}%
  \BibitemOpen
  \bibfield  {author} {\bibinfo {author} {\bibfnamefont {J.}~\bibnamefont
  {Tucker}}\ and\ \bibinfo {author} {\bibfnamefont {M.}~\bibnamefont
  {Feldman}},\ }\href {\doibase 10.1103/RevModPhys.57.1055} {\bibfield
  {journal} {\bibinfo  {journal} {Rev. Mod. Phys.}\ }\textbf {\bibinfo {volume}
  {57}},\ \bibinfo {pages} {1055} (\bibinfo {year} {1985})}\BibitemShut
  {NoStop}%
\bibitem [{\citenamefont {Horstman}\ and\ \citenamefont
  {Wolter}(1981)}]{horstman1981-gen}%
  \BibitemOpen
  \bibfield  {author} {\bibinfo {author} {\bibfnamefont {R.~E.}\ \bibnamefont
  {Horstman}}\ and\ \bibinfo {author} {\bibfnamefont {J.}~\bibnamefont
  {Wolter}},\ }\href@noop {} {\bibfield  {journal} {\bibinfo  {journal} {Phys.
  Lett. A}\ }\textbf {\bibinfo {volume} {82}},\ \bibinfo {pages} {43} (\bibinfo
  {year} {1981})}\BibitemShut {NoStop}%
\bibitem [{\citenamefont {Mooij}\ and\ \citenamefont
  {Klapwijk}(1983)}]{mooij1983-nem}%
  \BibitemOpen
  \bibfield  {author} {\bibinfo {author} {\bibfnamefont {J.~E.}\ \bibnamefont
  {Mooij}}\ and\ \bibinfo {author} {\bibfnamefont {T.~M.}\ \bibnamefont
  {Klapwijk}},\ }\href {\doibase 10.1103/PhysRevB.27.3054} {\bibfield
  {journal} {\bibinfo  {journal} {Phys. Rev. B}\ }\textbf {\bibinfo {volume}
  {27}},\ \bibinfo {pages} {3054} (\bibinfo {year} {1983})}\BibitemShut
  {NoStop}%
\bibitem [{\citenamefont {Maki}\ and\ \citenamefont
  {Tsuneto}(1964)}]{maki1964-pps1}%
  \BibitemOpen
  \bibfield  {author} {\bibinfo {author} {\bibfnamefont {K.}~\bibnamefont
  {Maki}}\ and\ \bibinfo {author} {\bibfnamefont {T.}~\bibnamefont {Tsuneto}},\
  }\href {\doibase 10.1143/PTP.31.945} {\bibfield  {journal} {\bibinfo
  {journal} {Prog. Theor. Phys.}\ }\textbf {\bibinfo {volume} {31}},\ \bibinfo
  {pages} {945} (\bibinfo {year} {1964})}\BibitemShut {NoStop}%
\bibitem [{\citenamefont {Maki}(1964)}]{maki1964-pps}%
  \BibitemOpen
  \bibfield  {author} {\bibinfo {author} {\bibfnamefont {K.}~\bibnamefont
  {Maki}},\ }\href {\doibase 10.1143/PTP.32.29} {\bibfield  {journal} {\bibinfo
   {journal} {Prog. Theor. Phys.}\ }\textbf {\bibinfo {volume} {32}},\ \bibinfo
  {pages} {29} (\bibinfo {year} {1964})}\BibitemShut {NoStop}%
\bibitem [{\citenamefont {Eliashberg}(1972)}]{eliashberg1972-iec}%
  \BibitemOpen
  \bibfield  {author} {\bibinfo {author} {\bibfnamefont {G.~M.}\ \bibnamefont
  {Eliashberg}},\ }\href@noop {} {\bibfield  {journal} {\bibinfo  {journal}
  {Sov. Phys. JETP}\ }\textbf {\bibinfo {volume} {34}},\ \bibinfo {pages} {668}
  (\bibinfo {year} {1972})}\BibitemShut {NoStop}%
\bibitem [{\citenamefont {Knoll}\ and\ \citenamefont
  {Keyes}(2003)}]{knoll2003-jnm}%
  \BibitemOpen
  \bibfield  {author} {\bibinfo {author} {\bibfnamefont {D.}~\bibnamefont
  {Knoll}}\ and\ \bibinfo {author} {\bibfnamefont {D.}~\bibnamefont {Keyes}},\
  }\href {\doibase 10.1016/j.jcp.2003.08.010} {\bibfield  {journal} {\bibinfo
  {journal} {J. Comp. Phys.}\ }\textbf {\bibinfo {volume} {193}},\ \bibinfo
  {pages} {357} (\bibinfo {year} {2003})}\BibitemShut {NoStop}%
\end{thebibliography}%

\end{document}